\documentstyle[11pt,aaspp4]{article}

\input epsf

\def\ni{\noindent}

\def\bar{\overline}

\def\bbE{\bar {\bf E}}

\def\beq{\begin{equation}}
\def\ee{\end{equation}}
\def\lsim{\mathrel{\rlap{\lower4pt\hbox{\hskip1pt$\sim$}}
    \raise1pt\hbox{$<$}}}
\def\gsim{\mathrel{\rlap{\lower4pt\hbox{\hskip1pt$\sim$}}
    \raise1pt\hbox{$>$}}}

\def\bB{\overline B}
\def\ts{\times}

\def\bbB{\overline {\bf B}}

\begin{document}

\setcounter{equation}{0}

\title{Pulsars With Jets May Harbor Dynamically Important Accretion Disks}

\medskip

\author{Eric G. Blackman \altaffilmark{1} and Rosalba Perna \altaffilmark{2}}
\affil{1. Department of Physics \& Astronomy,  
%Laboratory for Laser Energetics, 
University of Rochester, Rochester NY 14627}
\affil{2.  Department of Astrophysical Sciences, Princeton University,
Princeton, NJ 08544-0001}

%\documentstyle[11pt,aaspp4,flushrt]{article}
%\documentstyl[11pt]{article}
%\documentstyle[12pt,aasms4]{article}

\def\beq{\begin{equation}}
\def\eeq{\end{equation}}
\def\Mout{$\dot{M}_{\rm out}$}
\def\Min{$\dot{M}_{\rm in}$}
\def\Rout{${R}_{o}$}
\def\Rin{${R}_{i}$}

\centerline{(ApJ Lett., in press)}

\begin{abstract}
For many astrophysical sources with jets,
there is evidence for the contemporaneous presence of disks.
In contrast, pulsars such as the Crab and Vela show jets but
have not yet revealed direct evidence for accretion disks. 
Here we show that for such pulsars, an  
accretion disk radiating below detectable thresholds
may simultaneously account for  (1) observed  
deviations in the braking indices from 
that of the simple dipole, (2) observed pulsar timing ages, and 
(3) possibly even the jet morphology via a disk outflow
that interacts with the pulsar wind within, collimating and/or redirecting it.
\end{abstract}

\ni {\bf Key Words}: (stars:) pulsars: general; (stars:) 
pulsars: individual (Crab, Vela); accretion, accretion disks; 
ISM: jets and outflows; X-rays: ISM;  plasmas

\section{Introduction}

Recent X-ray observations have confirmed that collimated jets
can be found emanating from pulsar engines (Weisskopf et al. 2000;
Helfand et al. 2001; Gaensler et al. 2002; Slane 2002; Pavlov et
al. 2003).  These jets morphologically resemble those of young stellar
objects (Reipurth \& Bally 2001), planetary nebulae, (Balick \& Frank
2002) active galactic nuclei (Ferrari 1998) and microquasars (Mirabel
2001).  Jets and disks are likely associated in these systems,
even if the role of the disk vs. central object is not fully
understood with respect to jet production. Here we explore the
possibility that there may also be disks around jetted pulsars, and
that winds from these disks might 
collimate  the pulsar wind.  The motivation is
that relativistic pulsar winds do not seem to self-collimate easily
(e.g. Lyubarsky \& Eichler 2001; Bogovalov \& Khangoulian 2003;
Komissarov \& Lyubarsky 2003): strong outward electric forces compete
with the self-collimating collimating hoop stresses. 
Tsinganos \& Bogovalov (2002) simulated the
generic efficacy of relativistic wind collimation by an ambient
non-relativistic wind, but did not dynamically  relate the
external wind to the properties of an underlying disk.

Michel \& Dessler (1981) suggested that disks
formed from the stellar collapse offer a physical similarity
between radio and X-ray pulsars.
Such disks can also influence pulsar braking indices and timing ages:
the former often
deviate from those expected if spin-down were due only to dipole
radiation, and the latter often differ from inferred actual ages.
Various intrinsic mechanisms for these deviations have been proposed
(Macy 1974; Manchester \& Taylor 1977; Chanmugam \& Sang 1989; Beskin
et al. (1993); Chubarian et al. 2000; Wu et al. 2003), 
including an external torque
provided by a fallback accretion disk (e.g. Menou et al. 2001a;
Marsden et al. 2001; Alpar et al. 2001; Qiao et al. 2003; Shi \& Xu
2003). This would imply a weaker pulsar magnetic field  than
that inferred from the measured period $P$ and its derivative
$\dot{P}$. This has  been confirmed for one neutron star where the
magnetic field was measured from cyclotron lines (Bignami et
al. 2003).

Here we calculate the effect of a wind-emitting accretion
disk on the pulsar spin evolution by coupling the latter 
to evolution
equations for accretion and radius dependent mass loss. We constrain
the pulsar magnetic field strength and the 
initial outer accretion rate (or disk mass) 
for which the solutions match the observed periods, timing ages, and
braking indices of Crab and Vela.  The resulting solutions produce
sufficient disk-wind momenta to potentially collimate 
and help illuminate the pulsar winds.  
We describe our wind and accretion parameterization in section
2.  In Sec. 3 we show that the launch mechanism determines the
radial dependence of the mass outflow rate.  In Sec. 4 we calculate
the spin evolution of the pulsar and find best fit solutions for Crab
and Vela.  We compare the pulsar and disk-wind momenta in Sec. 5,
and conclude in Sec. 6.

\section{The accretion and disk-wind}

Given the  accretion rate at the outer disk radius ${\dot M}_{\rm out}$,
the time-dependent mass loss  from
a disk-wind launched between radii $R_{\rm out}$ and  $R< R_{\rm out}$ 
from the pulsar can be parameterized as 
\beq
\dot{M}_{\rm dw}(R,t) = \dot{M}_{\rm out}(t)\left[1-\left(
\frac{R}{R_{\rm out}(t)}\right)^p \right]\;.
\label{eq:Mwind}
\eeq
(Values of ${\dot M}_{\rm out}(t=t_0)$
and the parameter  $0< p <1$ will be discussed below.)
The accretion rate at the inner edge of the disk 
where the torque is exerted, is given by  
$\dot{M}_{\rm in}(t) = \dot{M}_{\rm out}(t) - \dot{M}_{\rm dw}(R_{\rm in},t)$.
The disk is allowed to penetrate the magnetosphere up to the magnetospheric
radius\footnote{
The surface where the pulsar
field develops a significant toroidal component (facilitating a 
Poynting flux outflow) likely has
an hourglass shape, being pinched at the disk midplane.
Field lines extending from the pulsar to large distances 
would develop toroidal fields closer to the light cylinder,  
due to the rapid density drop away from the disk.}
$R_{\rm in}\simeq 
2.55\times 10^8 \dot{M}_{\rm in,16}^{-2/7}M_{\rm NS,1}^{-1/7}
B_{12}^{4/7} {\rm cm}$ ($B_{12}10^{12}$ G is
the pulsar's magnetic field and $M_{\rm NS,1}M_\odot$ its mass)
where the torque exerted by the magnetic field on the disk
is of order the viscous torque.

To constrain ${\dot M}_{\rm out}(t)$, we appeal to previous work.
Chevalier (1989) studied fallback accretion, but without angular
momentum. Here we adopt the Menou et al. (2001a; see also Chatterjee
et al. 2000) analogy between fallback accretion and accretion of the
torus of gas formed by a disrupted star (Cannizzo et al. 1990).
Cannizzo et al. (1990) showed that there is a transient spreading
phase of duration $t_0$ during which the accretion is nearly constant,
$\dot{M}_{\rm out}(t)\sim \dot{M}_{\rm out}(t_0)$ but after which the
accretion rate declines as a power-law with time, $\dot{M}_{\rm
out}(t)= \dot{M}_{\rm out}(t_0)[t/t_0]^{-\delta}$, and the initial
accretion rate depends linearly on the product of the viscosity
parameter $\alpha_{ss}$ (Shakura \& Sunyaev 1973) and the initial mass
of the disk.  They also found that $\delta=19/16$ provides a good fit
to the late time evolution and that the outer edge of the disk then
evolves as $R_{\rm out} \propto (t/t_0)^{3/8}$.  The initial $R_{\rm
out}$ is determined by the angular momentum of the presupernova star
and the amount of ejected material.  Menou et al. (2001b) estimate
$R_{\rm out}(t_0) \approx 10^6-10^8$ cm for the typical angular
momentum in the presupernova star found in numerical simulations
(Heger, Langer \& Woosley 2000).  The time $t_0$ satisfies 
$
t_0\simeq 300 \left({\alpha_{ss}\over 0.1}\right)^{-1} \left({H_{\rm
out}/R_{\rm out}\over 0.1}\right)^{-2} \left({R_{\rm out}(t_0)\over
10^7{\rm cm}}\right)^{3/2} \left(M_{NS}\over M_\odot\right)^{-1/2}
{\rm sec},
$
%\label{eq:t0}
the viscous infall time at $R_{\rm out}(t_0)$, where $H_{\rm out}$ is the outer
disk height.

From (\ref{eq:Mwind}),
we can also write the total mechanical disk-wind luminosity as
\beq
L_{\rm dw}(t) = {1\over 2}
\int_{R_{\rm in}}^{R_{\rm out}} {d\over dR}
(\dot{M}_{\rm dw}(R,t) V_{\rm dw}^2) dR
\simeq \frac{GM_{\rm NS}
\left(V_{\rm dw}(R_{in})\over V_{\rm esc}\right)^{2}
\dot{M}_{\rm out}(t)\left[1-\left({R_{\rm in}\over R_{\rm out}}\right)^p\right]\;.}
{R_{\rm in}(t)}
\label{eq:Lwind}
\eeq
where  $V_{\rm esc}$ is the escape speed at the inner radius, and 
the last step holds when the total mass inside radius $R$
satisfies $M_{\rm tot}(R,t)\le M_{\rm NS} R/R_{\rm in}$. Note that the initial total 
disk mass is $\sim{\dot M}_{\rm out}(t_0)t_{0}$,
so our equations could be scaled in terms of this quantity.

\section{How $p$ can be constrained from disk-wind theory}

For magnetically mediated outflows (e.g. Spruit 1996), the mechanical wind luminosity 
is powered by Poynting flux from large scale fields.
The associated magnetic luminosity is  
\beq
L_{\rm dw}={c\over 4\pi}\int^{R_{\rm out}}_{R_{\rm in}} ({\bbE\ts \bbB})2\pi R dR=
{1\over 2}\int^{R_{\rm out}}_{R_{\rm in}} {V}_\phi 
{\bB}_{z} \bB_{\phi} R dR,
\label{poyn}
\eeq
where $\bbE$ is the mean surface electric field,
$\bB_z$ and   $\bB_\phi$ are the surface components
of the mean magnetic field, and $V_\phi$ is the Keplerian speed.
The surface magnetic fields in (\ref{poyn}) can be related to
the fields in the disk by using the volume integral
$\int \nabla\cdot\bbB dV=0$ taken over a 
wedge in the disk and converting to a surface integral.  
Using this, Tan \& Blackman (2003) showed that 
if  large scale disk fields are obtained from a helical dynamo, 
driven by disk turbulence and shear,  the 
surface magnetic stress  takes the form 
$
{\bB}_{z} \bB_{\phi}\sim 4\pi \rho  c_s^2  \alpha^{3/2}_{ss}(H/R),
\label{tor}
$
where $c_s$ is the mid-plane sound speed, $H$ is the disk height
and $\rho$ is the mid-plane density.
(The proportionality to $4\pi \rho c_s^2$ would arise
for any helical or non-helical dynamo, 
but the $\alpha_{ss}^{3/2}(H/R)$ factor would be replaced
by $\alpha_{ss}$ if we used  the averaged turbulent magnetic stress rather than
the magnetic stress associated with the large scale fields.)
Combining this magnetic stress 
with $\rho={{\dot M}_{\rm in}\over 4\pi R H V_r}$,
the equation for ${\dot M}_{\rm in}(t)$, the disk scalings (e.g. Frank et al. 2002) 
$V_r\simeq \alpha_{ss}c_s {H \over R}\simeq 
{\alpha_{ss} V_\phi }{H^2 \over R^2}$, 
and plugging the result into (\ref{poyn}) gives
\beq
{1\over 2}\int^{R_{\rm out}}_{R_{\rm in}} 
{\alpha_{ss}^{1/2}{\dot M}_{\rm in}V_\phi^2 \over R}dR = 
{\alpha^{1/2}_{ss}GM_{NS}{\dot M}_{\rm out}\over 2 R_i (R_{\rm out}/R_{\rm in})^p}
\int^{\lambda_o}_{1}\lambda^{p-2}d\lambda. 
\label{12}
\eeq
The last equation follows from 
the expression for $M_{\rm in}(t)$, $V_\phi= (GM_{NS}/R)^{1/2}$, and defining
$\lambda\equiv R/R_{\rm in}$ and $\lambda_o\equiv R_{\rm out}/R_{\rm in}$.
We can set (\ref{12}) equal to (\ref{eq:Lwind}), to obtain  
\beq
(1-p)(\lambda_o^{p}-1)
=(1-p)\left({V_{\rm esc}\over V_{\rm dw}}\right)^{2}
{\alpha^{1/2}_{ss}\over 2}
\int^{\lambda_o}_{1}\lambda^{p-2}d\lambda
= 
\left({V_{\rm esc}\over V_{\rm dw}}\right)^{2}
{\alpha_{ss}^{1/2}\over 2}(1-\lambda_0^{p-1}).
\label{13}
\eeq
Eqn. (\ref{13}) shows
that $p$ is not independent of $V_{\rm dw}/V_{\rm esc}$
and $\alpha_{ss}$ once a launch mechanism is chosen. 
The above formalism involves the relation that the mechanical disk-wind
luminosity at a given radius is a constant fraction of the accretion 
luminosity, so Ref (\ref{12}) implies 
that both $L_{\rm dw}$ and the accretion luminosity increase with decreasing $p$
if all other quantities were fixed. 
This formalism breaks down when the value of $p$ is so 
low that $V_{\rm dw}\rightarrow  c$.

\section{Pulsar spin-down}

The pulsar dipole luminosity for a neutron star of angular speed $\Omega$
and radius $R_{NS}$  
is given by 
\beq
L_{\rm pul}(t)\simeq \frac{B_*^2\,\Omega^4(t)\,R_{\rm NS}^6}{6c^3}
\;, 
\label{eq:Lpuls}
\eeq
where $B_*\equiv B\sin\theta$, and 
$\theta$ is the angle between the magnetic and spin axes. 
We have not considered the time evolution of $\sin\theta$ here.
This was studied by Beskin et al. (1993) 
and could  be included in future generalizations.
The rotational energy evolution is determined by this 
dipole radiation and a model dependent accretion torque so that
\beq
\dot{E}=I\Omega\dot{\Omega} = - \beta \Omega^4 + 2\dot{M}_{\rm in}R_m^2\Omega
\Omega_{\rm K}(R_m)\left[1-\frac{\Omega}{\Omega_{\rm K}(R_m)}\right]^{\gamma}\;,
\label{eq:spindown}
\eeq
where $I$ is the moment of inertia of the star, $\beta\equiv 
B^2\sin\theta^2 R_{\rm NS}^6/6c^3$, and we use a $\gamma=1$ model.
We also define the braking index and timing age as  
$ n(t)\equiv \frac{\Omega\ddot{\Omega}}{\dot{\Omega}^2}$
and $T(t)\equiv \frac{\Omega}{2\dot{\Omega}}$ respectively.

We determine the spin evolution of the star 
by numerically solving
Eq. (\ref{eq:spindown}) combined with the equations in section 2.
To do so, we also need the
pulsar's magnetic field, the initial spin period, and the initial  
accretion rate in the disk.  If the star were spinning down only by dipole
radiation, then its magnetic field would be directly determined from 
measuring  $P$ and $\dot{P}$. But when a disk
torque contributes to the spin-down, we only know
that the magnetic field strength must be smaller
than the value obtained for a pure dipole.  For young pulsars, we
expect $10^{11}\la B\la 10^{13}$ G.  The initial spin period $P_i$ is also
uncertain, but typically, $1 \la P_i\la 20$ msec. The initial accretion
rate is likewise unknown.  To constrain these parameters in our model,
we use observational values of the period, timing age and braking
index for the specific pulsar under consideration.  We then solve the
coupled system of equations using a grid of values for $P_i$, ${\dot
M}_{\rm out}(t_0)$ and $B_*$ to obtain a best fit parameter range that
reproduces the observed values of $P$, $T$ and $n$ to within a few
percent.

We specifically consider the Crab and Vela pulsars.  In 1972,  the
Crab's true age was 918 yr, and its measured timing age was $T=1243$~yr,
while the braking index was $n=2.51$ with period $P=33.1$ ms. Vela has a 
braking index of $n=1.4\pm 0.2$ (Lyne et 
al. 1996), a period $P=89$ ms, and a timing age $T=11.3$ kyr, for a
pulsar age (determined from the associated supernova remnant;
Aschenbach, Egger \& Trumper 1995) in the range 18-30 kyr. Here we 
use $\sim 25$ kyr.

Fig. 1a shows two solutions ($p=0.5$ and $p=0.01$) for the Crab, which
reproduce the measured values of $P$, $T$ and $n$.  The $p=0.5$
solution was obtained with $B_*=1.5\times 10^{12}$ G, $P_i=17.7$ msec,
and ${\dot M}_{\rm out}(t_0)=10^{29}$ g/s, while that with $p=0.01$
has $B_*=1.7\times 10^{12}$ G, $P_i=17.4$ msec, and ${\dot M}_{\rm
out}(t_0)=9.1\times 10^{27}$ g/s.  The bottom panel displays the
pulsar and wind luminosities [Eq.(\ref{eq:Lpuls}) and
Eq.(\ref{eq:Lwind})] during the pulsar lifetime.  Fig. 1b shows, for
the same values of $p$, the solutions for Vela which reproduce its
observations within the measurement uncertainties.  The solution with
$p=0.5$ was obtained with $B_*=7.7\times 10^{11}$ G, $P_i=7.9$ msec,
and $\dot{M}_{\rm out}(t_0)=3\times 10^{30}$ g/s, while that with
$p=0.01$ has $B_*=7.6\times 10^{11}$ G, $P_i=7.9$ msec, and ${\dot
M}_{\rm out}(t_0)=4.7\times 10^{29}$ g/s\footnote{ Multiplying $t_0$
with ${\dot M}_{\rm out}(t_0)$ from our fits gives initial disk masses
$\sim 0.01-0.5 M_\odot$.  The values of ${\dot M}_{\rm out}(t_0)$ in
our formal solutions, like those in accretion models of gamma-ray
bursts, are highly super-Eddington. Such disks would be dense, hot,
and thick from radiation trapping (Di Matteo et al. 2002).  Here the
Cannizzo et al. (1990) similarity solutions may be inapplicable, but
this does not strongly affect the present day pulsar parameter
solutions.}  Also displayed are the pulsar and wind luminosities, and
the observed lower limit on the jet luminosity. Pavlov et al. (2003)
estimated an energy injection rate $\ga 8\times 10^{33}$ g/s for the
Vela jet.  In all of our solutions, the corotation radius is less than
the magnetospheric radius at all times, so the pulsars are always spun
down by the disk.  The precise value of $p$ is not crucial to obtain a
good solution for $P$, $T$ and $n$; what matters for the spin-down
torque is the value of $\dot{M}_{\rm in}$, and a different $p$ can be
compensated with a different $\dot{M}_{\rm out}$ to yield a similar
$\dot{M}_{\rm in}$ (and hence a similar evolutionary track). This is
why the two lines in the top three panels of Fig.1 overlap. Note that
the higher $p$ curves for $L_{\rm dw}$ are $above$ the lower $p$
curves because ${\dot M}_{\rm out}$ is larger for the larger $p$
cases, and this counteracts the dependence of $L_{\rm dw}$ on $p$
discussed below Eq. (\ref{13}).

The particular solutions in Fig. 1 are not unique; 
multiple combinations of the parameters $B$, $P_i$ and $\dot{M}(t_0)$
can reproduce the values of $P$, $T$ and $n$ within the
observational uncertainties. However, we have shown solutions that
also produce a disk-wind powerful enough to influence the jets
observed in these sources, as discussed in more detail in the next
section.  For the solutions of Fig. 1, 
$R_{\rm in}\sim 9\times 10^7$ cm and $R_{\rm out}\sim 6\times
10^9$ cm for Crab and $R_{\rm in}\sim 10^8$ cm and $R_{\rm out}\sim
10^{10}$ cm for Vela.  For the Crab, the predicted optical
emission from a even a face-on thin disk\footnote
{The present day disk is assumed to be optically thick, geometrically
thin, and both viscous dissipation and irradiation
by the pulsar are  included (Perna et al.  2000; Perna \& Hernquist
2000).}  would be below the optical detection (Sollerman 2003). For
Vela, the observed optical flux limits require either (i) a disk
almost edge-on to the line of sight (ii) disruption of the outer
parts of the disk (maybe through interaction with the supernova
remnant) or (iii) a geometrically thick and
optically thin disk, like an advection dominated accretion flow (ADAF)
(e.g. Esin et al. 1997).

\section{Effect of  the disk-wind on the jet}

Relativistic Poynting flux pulsar winds do not easily self-collimate
(e.g. Lyubarsky \& Eichler 2001); as strong outward electric forces
compete with the collimating magnetic hoop stress. 
(This problem would not arise in 
non-relativistic outflows, as the electric field $<<$ magnetic field.)
We suggest instead that pulsar jets may be collimated 
by a surrounding disk wind.
 
A necessary condition 
for the disk wind to collimate the pulsar wind
is that the ram pressure associated with the disk wind 
exceed  the magnetic pressure associated with the 
Poynting flux dominated 
pulsar wind on scales at or less than those where the collimation is observed.
Since  the jets appear on scales $>>$ 
than that of the disk, it is reasonable to 
make this comparison far outside the pulsar's 
light cylinder.
Magnetic energy dominates the particle energy
in the pulsar wind, so the energy outflow rate can  be estimated as 
$L_{\rm pul} \sim  c (B^2/8\pi) 4\pi r^2$, where $r$ is a spherical radius.
Comparing the 
two pressures amounts to 
comparing  $B^2/8\pi$ to ${\dot M}_{\rm dw}(t) V_{\rm dw}/8 \pi r^2$.
But $B^2 r^2 = L_{\rm pul}/2c$, so this amounts to comparing 
${\dot M}_{\rm dw} V_{\rm dw}=L_{\rm dw}/V_{\rm dw}$ with $L_{\rm pul}/c$, namely 
comparing  the rate of change of the disk-wind
momentum to that of the pulsar wind.

Fig. 2 shows precisely this comparison for Crab and Vela.  The
solutions indicate that a collimating influence of the disk-wind on
the pulsar wind is possible for the $p=0.5$ case for both pulsars.
That $p=0.5$ provides higher disk-wind momenta and higher $L_{\rm
dw}/L_{\rm pul}$ than $p=0.01$ results because the ${\dot M}_{\rm out}$
which best fits the spin parameters is larger for the large $p$ cases.
(For the same ${\dot M}_{\rm out}$, the large $p$ case would have a
lower $L_{\rm dw}$ and momentum than the low $p$ case, as discussed
below Eq. (\ref{13}).)  Although this effect dominates, a competing
effect is that $V_{\rm dw}(R)$ increases with decreasing $p$: for both
pulsars, Eq. (\ref{13}) for $\alpha_{ss}=0.1$ reveals that $p=0.5$ and
$p=0.01$ correspond to $V_{\rm dw}/V_{\rm esc}\sim 0.2$ and $V_{\rm
dw}/V_{\rm esc}\sim 2$ respectively.  Collimation also requires that the
vector sum of the pulsar wind and disk wind momenta incurs a
significant vertical component.  Even if the disk wind were spherical,
when its ram pressure exceeds that of the pulsar wind the pulsar wind
would incur some collimation.  However, since the disk wind is largely
non-relativistic, magnetic self-collimation of the disk wind is not as
problematic as that of the pulsar wind.  Self-collimation of the disk
wind would favor even more collimation of the pulsar wind.  A more
detailed model of the interacting outflows is required to further
develop these ideas. Note that Tsinganos \& Bogovalov (2002) have
found numerically that relativistic central outflows could in
principle be collimated by non-relativistic disk winds, also
motivating more detailed models.  A standard limitation of such
numerical calculations is the assumption of the disk field geometry as
a given initial condition.

At the interface of the two winds, significant dissipation 
can occur, and accelerated
particles can produce the observed radiation. 
In addition, bending of the collimated jet 
could result from mutual 
non-axisymmetry of coaxial winds, or from misaligned pulsar and disk 
wind symmetry axes. 
Pavlov et al. (2003) suggested that a hidden wind from within the
supernova remnant is  needed to account for Vela's bent jet.
 
\section{Conclusions}

Winds from fallback accretion disks 
below presently detectable limits may deposit enough momenta 
to collimate  and bend the Crab and Vela pulsar winds 
and account for their observed jets. 
These  disks can also account for the observed periods, 
braking indices, and timing ages of the Crab and Vela pulsars 
through the action of the disk torque on pulsar spin-down.
Our results motivate further efforts to detect
disks around jetted pulsars, 
and more detailed models of the nested interacting
winds. Collimating a relativistic 
central wind by a surrounding accretion disk-wind may also 
be a relevant paradigm for active galactic nuclei and gamma-ray bursts.

We thank L. Hernquist, K. Menou, and the referee for useful comments.

\vfill
\eject

\noindent {Fig. 1: Evolution of the period, 
braking index, timing age, dipole luminosity and disk-wind luminosity
for Crab and Vela.
The points mark the observed values, though no such constraints exist
for the Crab jet power. 
The age for Vela  is taken to be that of the associated supernova remnant. 
Boldface lines correspond to $p=0.5$ and thin lines
to $p=0.01$. Results are rather insensitive to the value of $p$.}
\vspace{-.1cm} \hbox to \hsize{ 
\hfill \epsfxsize9cm \epsffile{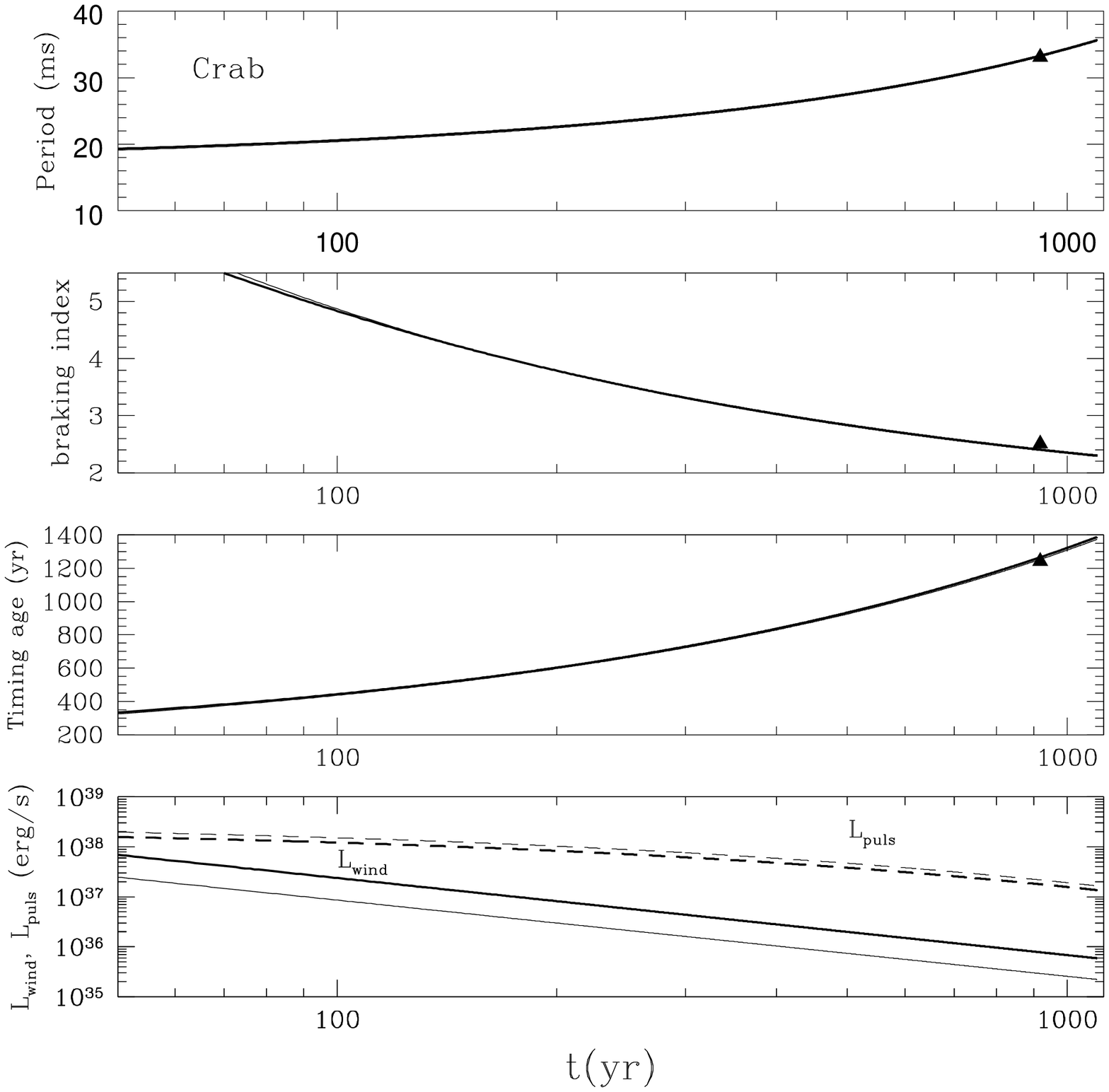} 
\hfill \epsfxsize9cm
\epsffile{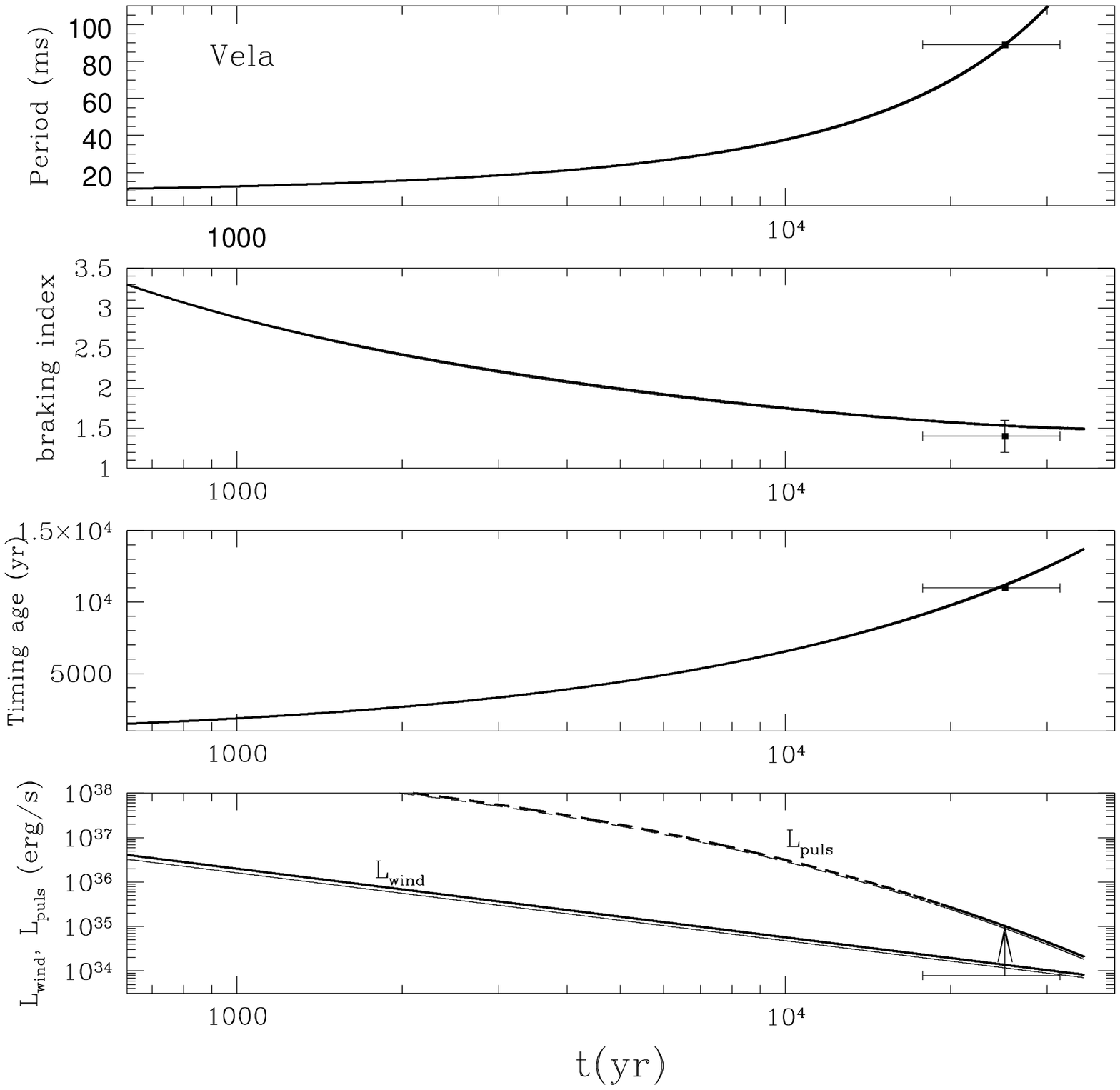} 
\hfill }

\bigskip
\bigskip
\bigskip
\bigskip
\bigskip
\bigskip
\bigskip

\vfill
\eject

\noindent{Fig. 2: Momentum per unit time ejected
by the disk-wind compared to that associated with the 
pulsar's Poynting flux. Boldface lines correspond to $p=0.5$ and thin lines
to $p=0.01$.} 
\vspace{-.1cm} \hbox to \hsize{ \hfill \epsfxsize10cm
\epsffile{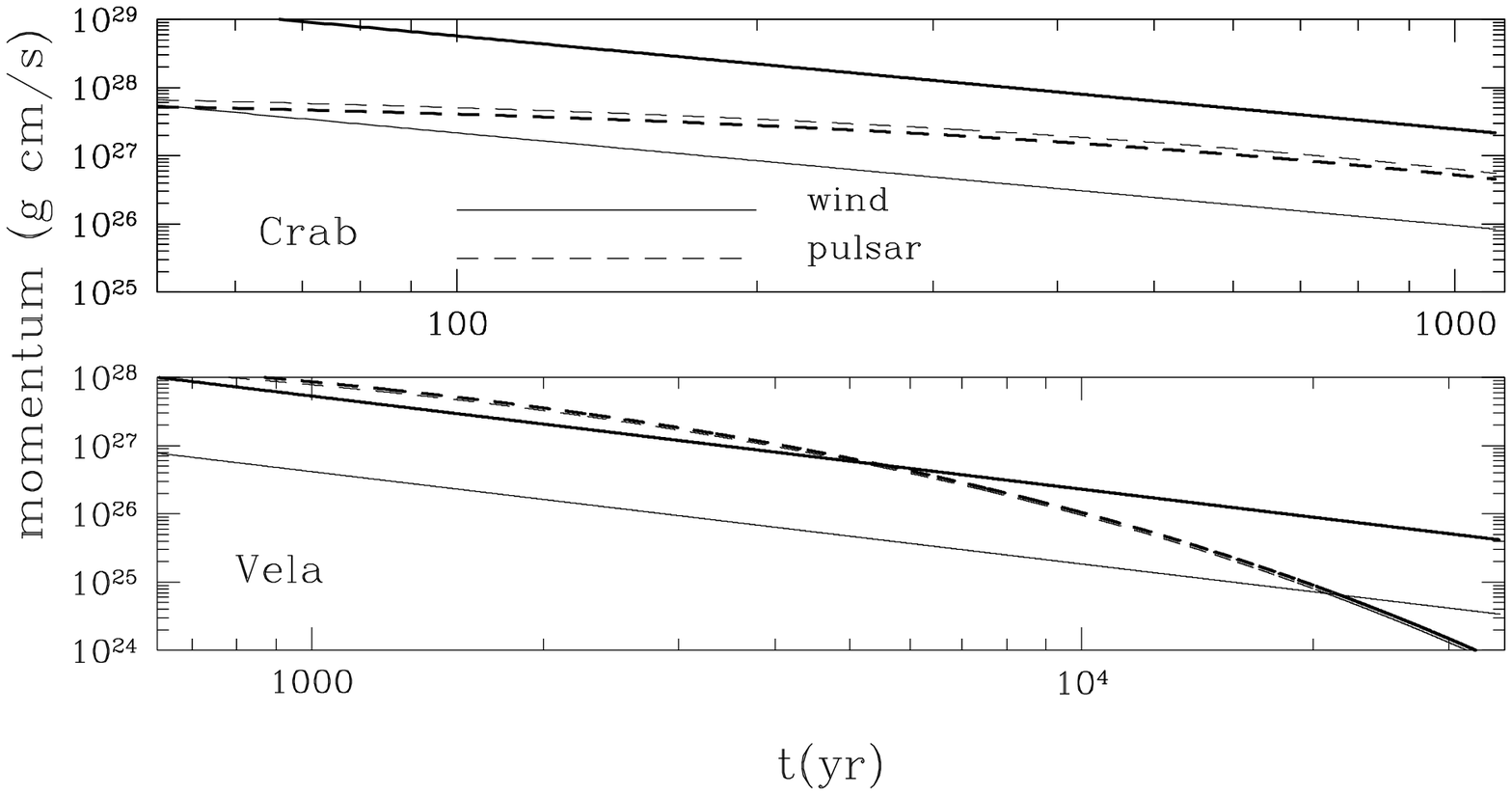} \hfill }


\begin{references}

\reference{} Alpar, M. A., Ankay, A., Yazgan, E. 2001, ApJ, 557L, 61
\reference{} Aschenbach, B., Egger, R., \& Trumper, J. 1995, Nature, 373, 587

\reference{} Balick, B.~\& Frank, A.\ 2002, ARAA, 40, 439 

\reference{} Bignami, G. F., Caraveo, P. A., De Luca, A., Mereghetti,
S. 2003, Nature, 423, 725
\reference{} Bogovalov, S. V.; Khangoulian, D. V.  2003, MNRAS, 336L, 53  

\reference{} Beskin V.S., Gurevich A.V., Istomin Ya. N, 1993 
``Physics of the Pulsar Magnetosphere'' (Cambridge Univ. Press: Cambridge)

\reference{} Cannizzo, J.K., Lee, H.M. \& Goodman, J. 1990, ApJ, 351,38

\reference{} Chanmugam, G.~\& Sang, Y.\ 1989, \mnras, 241, 295 

\reference{} Chatterjee, P., Hernquist, L. \& Narayan, R. 2000, ApJ 534, 373

\reference{} Chevalier, R.~A.\ 1989, ApJ, 346, 847

\reference{} Chubarian, E., Grigorian, H., 
Poghosyan, G., \& Blaschke, D.\ 2000, \aap, 357, 968 

\reference{} Di Matteo, T., Perna, R., Narayan, R. 2002, ApJ, 579, 706

\reference{} Esin A. A., McClintock J.E., Narayan R., 1997, ApJ, 489, 865 

\reference{} Ferrari, A.\ 1998, ARAA, 36, 539 

\reference{}Gaensler, B.~M., 
Arons, J., Kaspi, V.~M., Pivovaroff, M.~J., Kawai, N., \& Tamura, K.\ 2002, 
ApJ, 569, 878

\reference{} Heger, A.; Langer, N.; Woosley, S. E. 2000, ApJ, 528, 368

\reference{} Helfand, D.~J., 
Gotthelf, E.~V., \& Halpern, J.~P.\ 2001, ApJ, 556, 380

\reference{}Frank, J., King, 
A., \& Raine, D.~J.\ 2002, {\it Accretion Power in Astrophysics},  
(Cambridge University Press: Cambrdige UK.




\reference{} Komissarov, S. S. \& Lyubarsky, Y. E., preprint astro-ph/0306162
\reference{} Lyne, A.G., Pritchard, R.S., Graham-Smith, F. \& Camilo,
F. 1996, Nature, 381, 497
\reference{} Lyubarski, Y. \& Eichler, D. 2001, ApJ, 562, 494

\reference{} Macy, W.W., 1974, ApJ, 190, 153

\reference{} Manchester R.N. \& Taylor J.H., 1977, {\it Pulsars},
(Freeman: San Francisco).


\reference{} Marsden, D., Lingenfelter, R.E. \& Rothschild, R.E. 2001a, ApJL,
\reference{} Menou, K., Perna, R. \& Hernquist, L. 2001a, ApJ, 554L, 63
\reference{} Menou, K., Perna, R. \& Hernquist, L. 2001b, ApJ, 559, 1032

\reference{} Michel, F.~C.~\&  Dessler, A.~J.\ 1981, ApJ, 251, 654 


%\reference{} Michel, F.~C.~\& 
%Dessler, A.~J.\ 1983, Nature, 303, 48 

\reference{} Mirabel, I.~F.\ 2001, 
Astrophysics and Space Science Supplement, 276, 319 
\reference{} Pavlov, G. G.; Teter, M. A.; Kargaltsev, O.; Sanwal,
D. 2003, ApJ, 591, 1157

\reference{}Perna, R., Hernquist, L., \& Narayan, R. 2000, ApJ, 541, 344
\reference{} Perna, R. \& Hernquist, L. 2000, ApJ, 544, L57

\reference{}Reipurth, B.~\& 
Bally, J.\ 2001, ARAA, 39, 403 

\reference{} Sollerman, J. astro-ph/0306188

\reference{} Qiao, G. J., Xue, Y. Q, Xu, R. X., Wang, H. G., Xiao,
B. W., astro-ph/0306489
\reference{}Shakura, N.~I.~\&  Sunyaev, R.~A.\ 1973, A\&A, 24, 337 

\reference{} Shi, Y. \& Xu, R. X. 2003, ApJL in press, astro-ph/0307113

\reference{}Slane, P.\ 2002, ASP 
Conf.~Ser.~271: Neutron Stars in Supernova Remnants, 165 

\reference{} Spruit, H.~C.\ 1996, in NATO ASI
Proc.~477: {\it Evolutionary Processes in Binary Stars}, 
249, eds. R.A.M.J. Wijers, M.B. Davies and C.A. Tout, 
(Kluwer: Dordrecht) 

\reference{} Tsinganos K. \& Bogovalov S., 2002, MNRAS 337, 553.

\reference{} Tan, J. C. \& Blackman, E.G., 
2003, submited to ApJ, astro-ph/0307455

\reference{} Weisskopf, C., et al. 2000, ApJ, 536, L81

\reference{} Wu, F., Xu, R. X., Gil, J. 2003, A\&A, 409, 641

\end{references}
\end{document}